\documentclass[12pt,a4paper]{article}
\usepackage{amssymb,amsmath}
\usepackage[dvips]{lscape,graphicx}

\newcommand{\bc}{\begin{center}}
\newcommand{\ec}{\end{center}}
\newcommand{\bd}{\begin{displaymath}}
\newcommand{\ed}{\end{displaymath}}
\newcommand{\be}{\begin{equation}}
\newcommand{\ee}{\end{equation}}
\newcommand{\ba}{\begin{array}}
\newcommand{\ea}{\end{array}}
\newcommand{\bt}{\begin{tabular}}
\newcommand{\et}{\end{tabular}}

\newcommand{\ds}{\displaystyle}

\begin{document}

\title{Baryon asymmetry generation in the E$_6$CHM}

\author{R.~Nevzorov\footnote{On leave of absence from the Theory Department, SSC RF ITEP of NRC "Kurchatov Institute", Moscow, Russia.},
A.~W.~Thomas\qquad\qquad\\[5mm]
\itshape{ARC Centre of Excellence for Particle Physics at the Terascale and CSSM,}\\[0mm]
\itshape{Department of Physics, The University of Adelaide, Adelaide SA 5005, Australia}}

\date{}

\maketitle

\begin{abstract}{
\noindent
In the $E_6$ inspired composite Higgs model (E$_6$CHM) the strongly interacting sector
possesses an $SU(6)$ global symmetry which is expected to be broken down to its $SU(5)$ subgroup
at the scale $f\gtrsim 10\,\mbox{TeV}$. This breakdown results in a set of pseudo--Nambu--Goldstone
bosons (pNGBs) that includes one Standard Model (SM) singlet scalar,
a SM--like Higgs doublet and an $SU(3)_C$ triplet of scalar
fields, $T$. In the E$_6$CHM the $Z^L_{2}$ symmetry, which is a discrete subgroup of the $U(1)_L$ associated
with lepton number conservation, can be used to forbid operators which lead to rapid proton decay.
The remaining baryon number violating operators are sufficiently strongly suppressed because of the
large value of the scale $f$. We argue that in this variant of the E$_6$CHM a sizeable baryon number
asymmetry can be induced if CP is violated.
At the same time, the presence of the $SU(3)_C$ scalar triplet with mass in the few TeV range
may give rise to spectacular new physics signals that may be detected at the LHC in the near future.
}
\end{abstract}

\newpage
\section{Introduction}
Although the new scalar particle discovered at the LHC in 2012 is consistent with the Standard Model (SM) Higgs boson,
it could in principle be composed of more fundamental degrees of freedom. The idea of a composite Higgs boson was
proposed in the 70's \cite{Terazawa:1976xx} and 80's \cite{composite-higgs}. It implies the presence of a
strongly interacting sector in which electroweak (EW) symmetry breaking (EWSB) is generated dynamically.
Generically, in models of this type the composite Higgs tends to have a large quartic coupling $\lambda\gtrsim 1$.
At the same time, the observed SM-like Higgs boson is relatively light and corresponds to $\lambda\simeq 0.13$.
This indicates that the discovered Higgs state could possibly be a  pseudo-Nambu-Goldstone boson (pNGB)
originating from the spontaneous breakdown of an approximate global symmetry of the strongly interacting sector.

The minimal composite Higgs model (MCHM) \cite{Agashe:2004rs} contains two sectors (for a review, see                 Ref.~\cite{Bellazzini:2014yua}).
One of them involves weakly-coupled elementary particles, including all SM gauge bosons and SM fermions.
The second, strongly interacting sector gives rise to a set of bound states that, in particular, include composite
partners of the elementary particles; that is, massive fields with quantum numbers of all SM particles.

The composite sector of the MCHM possesses a global $\mbox{SO(5)}\times U(1)_X$ symmetry which is broken
down at the scale $f$ to $SO(4)\times U(1)'_X \cong SU(2)_W\times SU(2)_R\times U(1)'_X $,
which in turn contains the
$\mbox{SU(2)}_W\times U(1)_Y$ gauge group as a subgroup. This breakdown results in a set of pNGB states
that form the Higgs doublet.
The global custodial symmetry $SU(2)_{cust} \subset SU(2)_W\times SU(2)_R$ \cite{Sikivie:1980hm}
protects the Peskin--Takeuchi $\hat{T}$ parameter \cite{Peskin:1991sw}, which is strongly constrained by experimental
data \cite{Marandella:2005wd}, against the contributions induced by the composite states. The contributions of these
new states to electroweak observables were examined in Refs. \cite{EWPOCHM}--\cite{Vignaroli:2012si}.
The implications of the composite Higgs models were also
considered for Higgs physics \cite{Bellazzini:2012tv}--\cite{Azatov:2013ura}, \cite{Mrazek:2011iu}--\cite{Pomarol:2012qf},
gauge coupling unification \cite{Gherghetta:2004sq}--\cite{Barnard:2014tla}, dark matter \cite{Frigerio:2011zg}, \cite{Frigerio:2012uc},
\cite{Barnard:2014tla}--\cite{Asano:2014wra} and collider phenomenology \cite{Pomarol:2008bh}--\cite{Bellazzini:2012tv},
\cite{Barbieri:2008zt},  \cite{Pomarol:2012qf},  \cite{Redi:2011zi}--\cite{Delaunay:2013pwa}.
Non--minimal composite Higgs models were studied in Refs. \cite{Frigerio:2011zg}, \cite{Mrazek:2011iu}--\cite{Frigerio:2012uc},
\cite{Barnard:2014tla}--\cite{Asano:2014wra}, \cite{Cacciapaglia:2014uja}.

In these models the elementary and composite states with the same quantum numbers mix, so that at low energies
those states associated with the SM fermions (bosons) are superpositions of the corresponding elementary
fermion (boson) states and their composite fermion (boson) partners. In this partial compositeness
framework \cite{Contino:2006nn}-\cite{Kaplan:1991dc} the SM fields couple to the composite states,
including the Higgs boson, with a strength which is proportional to the compositeness fraction of each
SM field. In this case the mass hierarchy in the quark and lepton sectors can be reproduced if the
compositeness fractions of the first and second generation fermions are rather small. This also leads to the
suppression of  off-diagonal flavor transitions, as well as modifications of the $W$ and $Z$ couplings
associated with these light quark and lepton fields \cite{Contino:2006nn}, \cite{Agashe:2004cp}.

Even though this partial
compositeness considerably reduces the contributions of composite states to dangerous flavour--changing processes,
this suppression is not sufficient to satisfy all constraints. Within the composite Higgs models the constraints
that stem from the off--diagonal flavour transitions in the quark and lepton sectors were explored in
Refs. \cite{Barbieri:2012tu}--\cite{Vignaroli:2012si}, \cite{Redi:2011zi}, \cite{Blanke:2008zb}--\cite{Barbieri:2012uh}
and \cite{Redi:2013pga}, \cite{Barbieri:2012uh}--\cite{Csaki:2008qq}, respectively. In particular, it was argued that
in the case when the matrices of effective Yukawa couplings in the strongly interacting
sector are structureless, i.e anarchic matrices, the scale $f$ has to be larger than $10\,\mbox{TeV}$
\cite{Barbieri:2012tu}--\cite{Csaki:2008zd}, \cite{Redi:2011zi}, \cite{Blanke:2008zb}, \cite{Agashe:2006iy}.
This bound can be considerably alleviated in composite Higgs models with flavour symmetries
\cite{Barbieri:2008zt}--\cite{Barbieri:2012tu}, \cite{Redi:2011zi}--\cite{Redi:2013pga}, \cite{Barbieri:2012uh}, \cite{Cacciapaglia:2007fw},
in which the constraints originating from the Kaon and $B$ systems can be fulfilled if $f\gtrsim 1\,\mbox{TeV}$.
For such low values of the scale $f$ adequate suppression of the baryon number violating operators and
the Majorana masses of the left--handed neutrinos can be attained provided global $U(1)_B$ and $U(1)_L$ symmetries,
which guarantee the conservation of the baryon and lepton numbers respectively, are imposed.

In this note we focus on an $E_6$ inspired composite Higgs model (E$_6$CHM) in which the strongly interacting
sector is invariant under the transformations of an $SU(6)\times U(1)_L$ global symmetry \cite{Nevzorov:2015sha}.
In the weakly--coupled sector $U(1)_L$ is broken down to a $Z^L_2$ discrete symmetry which stabilizes the proton.
Since the composite sector in the E$_6$CHM does not possess any flavour or custodial symmetry, $SU(6)$ is expected
to be broken down to $SU(5)$, which in turn contains the SM gauge group, at a sufficiently high scale, $f\gtrsim 10\,\mbox{TeV}$.
This breakdown of the $SU(6)$ symmetry gives rise to a set of pNGBs that involves the SM--like Higgs
doublet, scalar coloured triplet and a SM singlet boson. Because the scale $f$ is so high, all baryon number violating
operators are sufficiently strongly suppressed so that the existing experimental constraints are satisfied. Nevertheless,
we argue that in the E$_6$CHM, with explicitly broken $U(1)_B$ baryon symmetry, the observed matter–-antimatter asymmetry can be induced if CP is broken.
The pNGB scalar coloured triplet plays a key role in this process and leads to
a distinct signature that can be observed at the LHC.

The layout of this article is as follows. In the next Section we discuss the E$_6$CHM with broken baryon symmetry.
In Section 3 we consider the process through which the baryon asymmetry is generated
and present our estimate of its value.
Section 4 concludes the paper.

\section{E$_6$CHM with baryon number violation}
The gauge and global symmetries of the E$_6$CHM, as well as its particle content, can originate from
a Grand Unified Theory (GUT)
based on the $E_6\times G_0$ gauge group. At some high energy scale, $M_X$,
the $E_6\times G_0$ gauge symmetry is broken
down to the $SU(3)_C\times SU(2)_W\times U(1)_Y \times G$ subgroup.
The gauge groups $G_0$ and $G$ are associated with the
strongly interacting sector, whereas $SU(3)_C\times SU(2)_W\times U(1)_Y$ is the SM gauge group.
Multiplets from the strongly coupled
sector are charged under both the $E_6$ and $G_0$ ($G$) gauge symmetries.
The weakly--coupled sector comprises fields that
participate in the $E_6$ interactions only. It is expected that in this sector different
multiplets of the elementary quarks and leptons
come from different fundamental $27$-dimensional representations of $E_6$.
All other components of these $27$--plets acquire
masses of the order of $M_X$. The corresponding splitting of the $27$--plets can occur
within a six--dimensional orbifold GUT model with
$N=1$ supersymmetry (SUSY) \cite{Nevzorov:2015sha} in which SUSY is broken somewhat
below the GUT scale $M_X$\footnote{Different
phenomenological aspects of the $E_6$ inspired models with low-scale SUSY breaking were
recently explored in \cite{e6ssm}-\cite{King:2008qb}.}.

In this orbifold GUT model the elementary quarks and leptons are components
of different bulk $27$--plets, while all fields from the strongly
interacting sector are localised on the brane, where the $E_6$ symmetry is broken down
to the $SU(6)\times SU(2)_N$ subgroup.
In the model under consideration $E_6$ is broken down to the SM gauge group
and $SU(2)_N$ symmetry is entirely broken.
Furthermore,  $SU(6)$ can remain an approximate global symmetry of the strongly coupled sector.
We assume that around the scale
$f\gtrsim 10\,\mbox{TeV}$ the $SU(6)$ global symmetry is broken down to $SU(5)$.
That, in turn,  contains the SM gauge group as a subgroup,
leading to a set of pNGB states which includes the SM--like Higgs doublet.

In the E$_6$CHM the $U(1)_L$ global symmetry, which ensures the conservation of lepton number,
can be used to suppress the operators
in the strongly interacting sector that may induce too large Majorana masses of the left--handed neutrino.
In the weakly--coupled elementary
sector this symmetry is supposed to be broken down to
\begin{equation}
Z^L_{2}=(-1)^{L} \, ,
\label{1}
\end{equation}
where $L$ is a lepton number, to guarantee that the left--handed neutrinos gain non-zero Majorana masses.
If $Z^L_2$ is an almost exact
discrete symmetry it also forbids potentially dangerous operators that give rise to rapid proton decay.
All other baryon number violating
operators in the model under consideration are sufficiently strongly suppressed
by the relatively large value of the scale $f$, as well as the
rather small mixing between elementary states and their composite partners.
Indeed, in the SM the effective operators responsible for
$\Delta B=2$ and $\Delta L=0$ are given by
\begin{equation}
\mathcal{L}_{\Delta B=2}=\frac{1}{\Lambda^5}\Biggl[ q_i q_j q_k q_m (d^c_n d^c_l)^{*} +
u^c_i d^c_j d^c_k u^c_m d^c_n d^c_l \Biggr]\,,
\label{2}
\end{equation}
where $q_i$ are doublets of the left-handed quarks, $u^c_i$ and $d^c_j$ are the right-handed
up- and down-type quarks and
the generation indices are $i,j,k,m,n,l=1,2,3$.

The $n-\bar{n}$ mixing mass can be deduced from this operator by simple dimensional analysis
to be $\delta m \simeq \varkappa \Lambda_{QCD}^6/ \Lambda^5$,
where $\varkappa$ is of order one and $\Lambda_{QCD}\simeq 200\,\mbox{MeV}$.
For $\Lambda \sim \mbox{few}\times 100\,\mbox{TeV}$ one finds the free $n-\bar{n}$ oscillation time
to be $\tau_{n-\bar{n}}\simeq 1/\delta m \simeq 10^8\,\mbox{s}$,
which is rather close to the present experimental limit \cite{Phillips:2014fgb}-\cite{Kronfeld:2013uoa}.
A similar bound on the scale $\Lambda$ comes from the rare nuclear decay searches looking for
the annihilation of the two nucleons $NN\to KK$, which
may be also induced by the operators (\ref{2}).
In this case one obtains a lower limit on $\Lambda$ of around $200-300\,\mbox{TeV}$.
On the other hand, in the composite Higgs models the small mixing between elementary states
and their composite partners leads to
$\Lambda\gtrsim \mbox{few}\times 100\,\mbox{TeV}$ when $f\gtrsim 10\,\mbox{TeV}$.

Thus, to ensure the phenomenological viability of the E$_6$CHM, the Lagrangian of
the strongly coupled sector of this model should respect
the $SU(6)\times U(1)_L$ global symmetry. Here we also assume that the low energy
effective Lagrangian of the E$_6$CHM
is invariant with respect to an approximate $Z^B_2$ symmetry,
which is a discrete subgroup of $U(1)_{B}$, i.e.
\begin{equation}
Z^B_{2}=(-1)^{3B} \, ,
\label{3}
\end{equation}
where $B$ is the baryon number. The $Z^B_2$ discrete symmetry does not forbid
baryon number violating operators (\ref{2}) but it does
provide an additional mechanism for the suppression of the proton decay.

In order to embed the E$_6$CHM into a Grand Unified Theory (GUT) based on the $E_6\times G_0$ gauge group,
the SM gauge couplings extrapolated to high energies using the renormalisation group equations (RGEs) should
converge to some common value near the scale $M_X$. Such an approximate unification of the SM gauge couplings
can be achieved if the right--handed top quark $t^c$ is entirely composite and the sector of weakly--coupled elementary
states involves  \cite{Nevzorov:2015sha}, \cite{Agashe:2005vg}
\begin{equation}
(q_i,\,d^c_i,\,\ell_i,\,e^c_i) + u^c_{\alpha} + \bar{q}+\bar{d^c}+\bar{\ell}+\bar{e^c} \, ,
\label{4}
\end{equation}
where $\alpha=1,2$ and $i=1,2,3$. Here we have denoted the left-handed lepton
doublet by $\ell_i$, the right-handed charged leptons
by $e_i^c$, while the extra exotic states in Eq.~(\ref{4}), $\bar{q},\,\bar{d^c},\,\bar{\ell}$
and $\bar{e^c}$, have exactly opposite
$SU(3)_C\times SU(2)_W\times U(1)_Y$ quantum numbers to the
left-handed quark doublets, right-handed down-type quarks,
left-handed lepton doublets and right-handed charged leptons, respectively. This scenario also implies
that the strongly coupled sector gives rise to the composite ${\bf 10} + {\bf \overline{5}}$ multiplets of $SU(5)$.
These multiplets get combined with $\bar{q},\,\bar{d^c},\,\bar{\ell}$ and $\bar{e^c}$,
resulting in a set of vector--like states.
The only exceptions are the components of the $10$--plet that correspond to $t^c$,
which survive down to the EW scale.

In the simplest case the composite ${\bf 10} + {\bf \overline{5}}$ multiplets of $SU(5)$ stem from one
${\bf{15}}$--plet and two ${\bf \overline{6}}$--plets (${\bf \overline{6}}_1$ and ${\bf \overline{6}}_2$) of $SU(6)$.
These $SU(6)$ representations have the following decomposition in terms of $SU(5)$ representations:
${\bf 15}={\bf 10} \oplus {\bf 5}$ and ${\bf \overline{6}}={\bf \overline{5}} \oplus {\bf 1}$.
The components of these ${\bf{15}}$, ${\bf \overline{6}}_1$ and ${\bf \overline{6}}_2$ multiplets decompose
under $SU(3)_C\times SU(2)_W\times U(1)_Y$ as follows:
\begin{equation}
\ba{ll}
\ba{rcl}
{\bf 15} &\to& Q = \left(3,\,2,\,\ds\frac{1}{6}\right)\,,\\[2mm]
&& t^c = \left(3^{*},\,1,\,-\ds\frac{2}{3}\right)\,,\\[2mm]
&& E^c = \Biggl(1,\,1,\,1\Biggr)\,,\\[2mm]
&& D = \left(3,\,1,\,-\ds\frac{1}{3} \right)\,,\\[2mm]
&& \overline{L}=\left(1,\,2,\,\ds\frac{1}{2}\right)\,;
\ea
\qquad
\noindent
\ba{rcl}
{\bf \overline{6}}_{\alpha} &\to & D^c_{\alpha} = \left(\bar{3},\,1,\,\ds\frac{1}{3} \right)\,,\\[2mm]
& & L_{\alpha} = \left(1,\,2,\,-\ds\frac{1}{2} \right)\,,\\[2mm]
& & N_{\alpha} = \Biggl(1,\,1,\,0 \Biggr)\,,
\ea
\ea
\label{5}
\end{equation}
where $\alpha=1,2$. The first and second quantities in brackets are the $SU(3)_C$ and $SU(2)_W$ representations,
while the third are the $U(1)_Y$ charges. The large mass of the top quark can be generated only if
$t^c$ is $Z^B_2$-odd. As a consequence all components of the ${\bf{15}}$--plet have to be odd under the
$Z^B_2$ symmetry. After the $SU(6)$ symmetry breaking a ${\bf 5}$--plet from the ${\bf{15}}$--plet
and ${\bf \overline{5}}$--plet from the ${\bf \overline{6}}_2$ form vector--like states. The corresponding mass terms
are allowed if all components of ${\bf \overline{6}}_2$ are $Z^B_2$-odd. In principle, the components of
${\bf \overline{6}}_1$ multiplet could be either even or odd under the $Z^B_2$ symmetry. Hereafter we assume
that $D^c_{1}$, $L_{1}$ and $N_{1}$ are $Z^B_2$--even.

The breakdown of the $SU(6)$
symmetry also induces the Majorana masses for the SM singlet states $N_1$ and $N_2$. The mixing of
these states is suppressed because of the approximate $Z^B_2$ symmetry.
As discussed above, the remaining components
of the $SU(6)$ multiplets ${\bf{15}}$ and ${\bf \overline{6}}_1$ get combined with $\bar{q},\,\bar{d^c},\,\bar{\ell}$
and $\bar{e^c}$ leading to the composite right--handed top quark $t^c$ and a set of vector--like states.
In general all extra exotic fermions tend to gain masses which are a few times larger than $f$. Therefore it is unlikely
that these states will be detected at the LHC in the near future. In the next section we consider
the phenomenological implications of this variant of the E$_6$CHM, assuming that $N_1$ is considerably
lighter than other exotic fermion states and has a mass which is somewhat smaller than $f$.

\section{Generation of baryon asymmetry}

As mentioned earlier, the breakdown of the $SU(6)$ to its $SU(5)$ subgroup gives rise to a set of pNGB states.
The masses of all pNGB states are expected to be considerably lower than $f\gtrsim 10\,\mbox{TeV}$, so that these
resonances are the lightest composite states. The corresponding set involves
eleven pNGB states. One of them, $\phi_0$, is a real SM singlet scalar. The collider signatures associated with the
presence of such a scalar, in the limit where CP is conserved, were examined in Ref.~\cite{Nevzorov:2016fxp}.
Ten other pNGB states form a fundamental representation of unbroken $SU(5)$, i.e. $\tilde{H}=\bf{5}$.
The first two components of $\tilde{H}$ transform as an $SU(2)_W$ doublet and
correspond to the SM--like Higgs doublet, $H$,
whereas three other components of $\tilde{H}$ are associated with the $SU(3)_C$ triplet of scalar fields $T$.
The pNGB effective potential $V_{eff}(\tilde{H}, T, \phi_0)$ is induced by the interactions
of the elementary states with their composite partners that explicitly violate the $SU(6)$ global symmetry. In the model
under consideration substantial tuning, $\sim 0.01\%$, is required to get $v\ll f$ and a $125\,\mbox{GeV}$ Higgs boson,
because the scale $f$ is so large. Nevertheless, it has been shown that in models similar to the E$_6$CHM there exists
a part of the parameter space where the $SU(2)_W\times U(1)_Y$ gauge symmetry is broken to $U(1)_{em}$,
whilst $SU(3)_C$ remains intact  \cite{Frigerio:2011zg}, \cite{Barnard:2014tla}. In these composite Higgs models
the $SU(3)_C$ triplet scalar, $T$, tends to be substantially heavier than the Higgs scalar.

In the interactions with other SM particles the Higgs boson manifests itself as a $Z^B_2$-even state.
Therefore all other pNGB states should be also even under the $Z^B_2$ symmetry.
The gauge and $Z_2^B$ symmetries
allow the decays of the scalar triplet $T$ into up and down antiquarks.
At the same time, the decays of the $SU(3)_C$ scalar
triplet into a charged lepton and an up quark as well as into a neutrino and a down quark
are forbidden by the almost exact $Z^L_2$ symmetry.
Since the fractions of compositeness of the first and second generation quarks are rather small, the decay mode
$T\to\bar{t}\bar{b}$ tends to be the dominant one. At the energies $E\lesssim f$ almost all resonances of the composite
sector, except the pNGB states, can be integrated out and all baryon number violating operators are strongly suppressed,
so that baryon number is conserved to a very good approximation. In this limit $T$ manifests itself in the interactions with
other SM bosons and fermions as a diquark, i.e. an $SU(3)_C$ scalar triplet with $B=-2/3$.

The presence of this exotic $SU(3)_C$ scalar triplet with mass $m_{T}$ in the few TeV range
makes possible the generation
of the baryon asymmetry via the out--of equilibrium decays of $N_1$, provided $N_1$
is the lightest exotic fermion in the spectrum.
Indeed, the Majorana mass of $N_1$ is set by $f$, while $m_{T}\ll f$.
As a result the decays $N_{1}\to T+\bar{d}_i$ and $N_{1}\to T^{*}+ d_i$ are kinematically allowed.
Since at low energies $E\lesssim f$ the $SU(3)_C$ scalar triplet, $T$, manifests itself in the interactions with
other SM states as a diquark, the Majorana fermion $N_1$ can decay into final states with
baryon numbers $\pm 1$. The interactions of $N_1$ and $N_2$ with the pNGB state $T$ and down-type
quarks are described by the Lagrangian
\begin{equation}
\mathcal{L}_{N}= \sum_{i=1}^3 \Biggl( g^{*}_{i1} T d^c_i N_1 + g^{*}_{i2} T d^c_i N_2 + h.c.\Biggr) \, .
\label{6}
\end{equation}
In the exact $Z^B_2$ symmetry limit, the couplings $g_{i1}$ have to vanish.
Therefore the approximate $Z^B_2$ symmetry
ensures that the $g_{i1}$ couplings are somewhat suppressed
in comparison with $g_{i2}$, i.e. $|g_{i1}| \ll |g_{i2}|$.

The process of the baryon asymmetry generation is controlled by the flavour
CP (decay) asymmetries $\varepsilon_{1,\, k}$
that appear on the right--hand side of Boltzmann equations.
There are three decay asymmetries associated with three quark flavours
$d,\,s$ and $b$. These are given by
\begin{equation}
\varepsilon_{1,\,k}=\dfrac{\Gamma_{N_1 d_{k}}-\Gamma_{N_1 \bar{d}_{k}}}
{\sum_{m} \left(\Gamma_{N_1 d_{m}}+\Gamma_{N_1 \bar{d}_{m}}\right)} \, ,
\label{7}
\end{equation}
where $\Gamma_{N_1 d_{k}}$ and $\Gamma_{N_1 \bar{d}_{k}}$ are partial decay widths of $N_1\to d_k + T^{*}$ and
$N_1\to \overline{d}_k + T$ with $k,m=1,2,3$.  At the tree level the CP asymmetries (\ref{7})
vanish because (see \cite{King:2008qb})
\begin{equation}
\Gamma_{N_1 d_{k}}=\Gamma_{N_1 \bar{d}_{k}}=\dfrac{3 |g_{k1}|^2}{32 \pi}\,M_1\,,
\label{8}
\end{equation}
where $M_1$ is the  Majorana mass of $N_1$. However, if CP invariance is broken the
non--zero contributions to the CP asymmetries
arise from the interference between the tree--level amplitudes of the $N_1$ decays and
the one--loop corrections to them.
The tree--level and one--loop diagrams that give contributions to the CP asymmetries associated with the decays
$N_1\to \overline{d}_k + T$ can be found in \cite{King:2008qb}. Assuming that the $SU(3)_C$ scalar triplet $T$
is much lighter than $N_1$ and $N_2$, the direct calculation of the appropriate one--loop diagrams
gives\footnote{These calculations are very similar to the ones performed in the case of thermal leptogenesis
\cite{CPasym-SM} (for the review see \cite{Davidson:2008bu})}
\begin{equation}
\begin{array}{rcl}
\varepsilon_{1,\,i}&=&\dfrac{1}{(8\pi)}\dfrac{1}{(\sum_{m=1}^3 |g_{m1}|^2)}\Biggl[
\sum_{n=1}^3 \mbox{Im}(g^{*}_{i1} g_{i2} g^{*}_{n1} g_{n2}) \sqrt{x} \Biggl(\dfrac{3}{2(1-x)}+1\\
&&-(1+x)\ln\dfrac{1+x}{x}\Biggr)+\sum_{n=1}^3 \mbox{Im}(g^{*}_{i1} g_{i2} g_{n1} g^{*}_{n2}) \dfrac{3}{2(1-x)}
\Biggr]\,,
\end{array}
\label{9}
\end{equation}
where $x=(M_2/M_1)^2$ and $M_2$ is the Majorana mass of $N_2$.

In order to calculate the total baryon asymmetries induced by the decays of $N_1$, the system of Boltzmann equations
that describe the evolution of baryon number densities have to be solved. The corresponding solution should be
somewhat similar to the solutions of the Boltzmann equations for thermal leptogenesis; so that in the first approximation
the generated baryon asymmetry can be estimated using an approximate formula
given in Ref.~\cite{Davidson:2008bu}\footnote{The
induced baryon asymmetry is partially converted into lepton asymmetry due to $(B+L)$--violating sphaleron interactions \cite{Kuzmin:1985mm}.
Here we ignore sphaleron processes.}
\begin{equation}
Y_{\Delta B}\sim 10^{-3}\biggl(\sum_{k=1}^3 \varepsilon_{1,\,k} \eta_k\biggr)\,,
\label{10}
\end{equation}
where $Y_{\Delta B}$ is the baryon asymmetry relative to the entropy density, i.e.
$$
Y_{\Delta B}=\dfrac{n_B-n_{\bar{B}}}{s}\biggl|_0=(8.75\pm 0.23)\times 10^{-11}\,.
$$
In Eq.~(\ref{10}) $\eta_k$ are efficiency factors. A thermal population of $N_1$ decaying completely
out of equilibrium without washout effects would lead to $\eta_{k}=1$. However washout processes
reduce the induced asymmetries by the factors $\eta_k$, where $\eta_k$ varies from 0 to 1.

To simplify our analysis we assume that the pNGB state $T$ couples primarily to
the $b$--quarks, i.e. $|g_{31}|\gg |g_{21}|,\, |g_{11}|$ and $|g_{32}|\gg |g_{22}|,\, |g_{12}|$,
and $N_1$ is substantially lighter than all other exotic fermions. In particular, we set $M_2=10\cdot M_1$.
The imposed hierarchical structure of the Yukawa couplings implies that the decay asymmetries
$\varepsilon_{1,\,2}$ and $\varepsilon_{1,\,1}$ are much smaller than $\varepsilon_{1,\,3}$
and can be neglected. If $g_{31}=|g_{31}| e^{i\varphi_{31}}$ and $g_{32}=|g_{32}| e^{i\varphi_{32}}$
then in the limit $x\gg 1$ one finds
\begin{equation}
\varepsilon_{1,\,3}\simeq -\dfrac{1}{(4\pi)}\dfrac{|g_{32}|^2}{\sqrt{x}}\sin 2\Delta\varphi\,,\qquad\qquad
\Delta\varphi=\varphi_{32}-\varphi_{31}\,.
\label{11}
\end{equation}
The CP asymmetry (\ref{11}) vanishes when all Yukawa couplings are real, i.e. CP invariance is preserved.
The decay asymmetry $\varepsilon_{1,\,3}$ attains its maximum absolute value when $\Delta\varphi=\pm \pi/4$,
i.e. $\sin 2\Delta\varphi$ is equal to $\pm 1$.

In order to estimate the efficiency factor $\eta_3$, we concentrate on the so--called strong
washout scenario (see, for example \cite{Davidson:2008bu}) for which
\begin{equation}
\begin{array}{c}
\eta_3 \simeq H(T=M_1)/\Gamma_{3}\,,\\[3mm]
\Gamma_3 = \Gamma_{N_1 d_{3}}+\Gamma_{N_1 \bar{d}_{3}}=\dfrac{3 |g_{31}|^2}{16 \pi}\,M_1\,,
\qquad\qquad
H=1.66 g_{*}^{1/2}\dfrac{T^2}{M_{Pl}}\,,
\end{array}
\label{12}
\end{equation}
where $H$ is the Hubble expansion rate and $g_{*}=n_b+\dfrac{7}{8}\,n_f$ is the number of relativistic
degrees of freedom in the thermal bath. Within the SM $g_{*}=106.75$, whereas in the E$_6$CHM
$g_{*}=113.75$ for $T\lesssim f$. Eqs.~(\ref{12}) indicate that $\eta_3$ increases with diminishing of
$|g_{31}|$. Thus this coupling of $N_1$ to the pNGB state $T$ can be adjusted so that $\eta_3$ becomes
relatively close to unity. In particular, from Eqs.~(\ref{12}) it follows that for $|g_{31}|\simeq 10^{-6}$
and $M_1\simeq 10\,\mbox{TeV}$ the efficiency factor $\eta_3$ is around $0.25$.

\begin{figure}
\includegraphics[width=75mm,height=55mm]{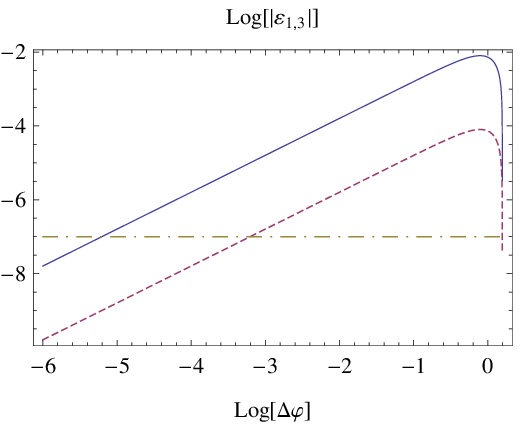}\qquad
\includegraphics[width=75mm,height=55mm]{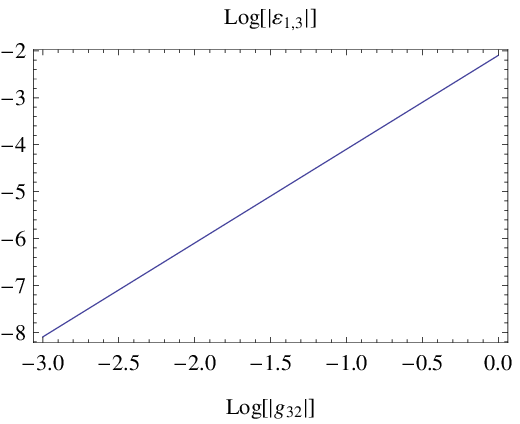}\\[2mm]
\hspace*{3.5cm}{\bf (a)}\hspace*{8cm}{\bf (b) }\\
\caption{Logarithm (base 10) of the absolute value of the CP asymmetry $\varepsilon_{1,\,3}$ for $g_{11}=g_{21}=g_{12}=g_{22}=0$
and $M_2=10\cdot M_1$. In {\it (a)} the absolute value of $\varepsilon_{1,\,3}$ is given as a function of
logarithm (base 10) of $\Delta \varphi$ for $|g_{32}|=1$ (solid line) and $|g_{32}|=0.1$ (dashed line).
In {\it (b)} the maximal absolute value of $\varepsilon_{1,\,3}$ is presented as a function of logarithm (base 10) of
$|g_{32}|$ for $\Delta \varphi=\pi/4$.
}
\label{fig1}
\end{figure}

If the efficiency factor $\eta_3$ is sufficiently large, i.e. $\eta_3 \sim 0.1-1$, the baryon asymmetry is determined by the
induced decay asymmetry $\varepsilon_{1,\,3}$. Indeed, from Eqs.~(\ref{9}) and (\ref{11}) one can see that in the
limit $g_{21}=g_{11}\to 0$ the CP asymmetries $\varepsilon_{1,\,2}$ and $\varepsilon_{1,\,1}$ vanish while
$\varepsilon_{1,\,3}$ does not depend on the absolute value of the Yukawa coupling $g_{31}$. Therefore,
for a given ratio $M_2/M_1$ the CP asymmetry $\varepsilon_{1,\,3}$ is set by $|g_{32}|$ and the combination of
the CP--violating phases $\Delta \varphi$. The dependence of the absolute value of $\varepsilon_{1,\,3}$ on these
parameters is examined in Fig.~\ref{fig1}, where we fix $(M_2/M_1)=10$. Since the Yukawa coupling of $N_2$ to
$SU(3)_C$ scalar triplet and $b$-quark is not suppressed by the $Z_2^B$ symmetry, $|g_{32}|$ is expected to be
relatively large, i.e. $|g_{32}| \gtrsim 0.1$. In Fig.~1a we plot the absolute value of $\varepsilon_{1,\,3}$ as a function
of $\Delta \varphi$ for $|g_{32}| = 0.1$ and $|g_{32}| = 1$. Fig.~1a illustrates that the CP asymmetry
$\varepsilon_{1,\,3}$ attains its maximum absolute value $\sim 10^{-4}-10^{-2}$ for $\Delta \varphi\simeq \pi/4$.
Thus a larger value of $|\varepsilon_{1,\,3}|$ can lead to a phenomenologically acceptable
baryon density only for sufficiently small
values of efficiency factor, $\eta_3=10^{-5}-10^{-3}$.
When this factor is reasonably large, i.e. $\eta_3 \sim 0.1-1$,
and $|g_{32}|\simeq 0.1$ a phenomenologically acceptable value of the baryon density,  corresponding to
$\varepsilon_{1,\,3}\lesssim10^{-7}-10^{-6}$, is generated  only if the combination of the CP--violating phases
$\Delta \varphi$ is rather small, i.e. $\Delta\varphi\lesssim 0.01$. This demonstrates that the appropriate baryon
asymmetry can be obtained within the E$_6$CHM even if CP is approximately preserved.

In Fig.~1b the dependence of the maximum value of $|\varepsilon_{1,\,3}|$ on $|g_{32}|$ is studied.
From Eq.~(\ref{11}) and Fig.~1b it follows that the maximum absolute value of this CP asymmetry
grows monotonically with increasing of $|g_{32}|$. Fig.~1b also indicates that the appropriate
baryon density associated with $\varepsilon_{1,\,3}\gtrsim 10^{-7}-10^{-6}$ can be
obtained even if the absolute value of the corresponding Yukawa coupling varies from $0.01$ to $0.1$.

\section{Conclusions}
In the $E_6$ inspired composite Higgs model (E$_6$CHM) the approximate $SU(6)$ global symmetry of
the strongly coupled sector is supposed to be broken down at the scale $f\gtrsim 10\,\mbox{TeV}$ to its
$SU(5)$ subgroup, which incorporates the $SU(3)_C\times SU(2)_W\times U(1)_Y$ gauge symmetry.
Within this model the operators that may result in rapid proton decay can be suppressed by
a $Z^L_{2}$ discrete symmetry. Since the scale $f$ is so large all baryon number violating operators,
which are not forbidden by the $Z^L_{2}$ symmetry, are sufficiently strongly suppressed.
Nonetheless, this variant of the E$_6$CHM leads to baryon number violating processes, like neutron-antineutron
oscillations, that are going to be searched for in future experiments \cite{Phillips:2014fgb}-\cite{Kronfeld:2013uoa}.
To ensure the approximate unification of the SM gauge couplings, that makes possible the embedding of the E$_6$CHM
into a suitable GUT, this model involves extra matter.
Additional matter multiplets give rise to a composite right-handed top quark
and a set of exotic fermions that, in particular, includes two SM singlet Majorana states $N_1$ and $N_2$.
In general all exotic fermions acquire masses which are somewhat larger than $f$.
In our analysis we assumed that $N_1$ is the lightest exotic fermion, with a mass around $10\,\mbox{TeV}$.

The pNGB states, which originate from the breakdown of $SU(6)$ to its $SU(5)$ subgroup, are the lightest
composite resonances in the E$_6$CHM. The corresponding set of states contains one SM singlet scalar,
a SM--like Higgs doublet and an $SU(3)_C$ triplet of scalar fields, $T$. The masses of all these resonances
tend to be substantially lower than $f$. At energies $E\lesssim f$ baryon number is preserved
to a very good approximation and the $SU(3)_C$ scalar triplet $T$ manifests itself in the interactions with
the SM particles as a diquark. We argued that in this variant of the E$_6$CHM the baryon asymmetry
can be generated via the out--of equilibrium decays of $N_1$ into final states with baryon numbers $\pm 1$,
i.e., $N_{1}\to T+\bar{d}_i$ and $N_{1}\to T^{*}+ d_i$, provided CP is violated. Moreover, if the absolute value of the
Yukawa coupling of $N_2$ to $T$ and $b$--quark varies in the range $0.1$ to $1$
a phenomenologically acceptable baryon
density may be obtained, even when all CP--violating phases are quite small ($\lesssim 0.01$). In this case the
approximate CP conservation leads to suppression of the electric dipole moments (EDMs) of the neutron, elementary
states and atoms that have not been observed in numerous experiments but can be measured
in the near future (see \cite{Kronfeld:2013uoa}).
Since the couplings of $N_1$, $N_2$ and $T$ to the first and second generation quarks are tiny, their contributions
to the baryon number violating processes, like $n-\bar{n}$ oscillations, are sufficiently strongly suppressed. On the
other hand, the $SU(3)_C$ scalar triplet $T$, with mass in the few TeV range, can be pair produced at the LHC and
predominantly decays into $T\to \bar{t}+\bar{b}$, leading to some enhancement of the cross section
of $pp\to t\bar{t}b\bar{b}$. Thus the scenario under consideration emphasizes the importance of the
complementarity of different experiments.

\section*{Acknowledgements}
RN is grateful to Z.~Berezhiani, D.~A.~Demir, R.~Erdem, H.~Fritzsch, J.~F.~Gunion, S.~F.~King,  A.~Kobakhidze, M.~M\"{u}hlleitner, X.~Tata and P.~Zerwas
for helpful discussions. This work was supported by the University of Adelaide and the
Australian Research Council through the ARC Center of Excellence
in Particle Physics at the Terascale (CE 110001004).


\newpage
\begin{thebibliography}{99}
\bibitem{Terazawa:1976xx}
H.~Terazawa, K.~Akama, Y.~Chikashige,
%``Unified Model of the Nambu-Jona-Lasinio
%Type for All Elementary Particle Forces,''
Phys.\ Rev.\ D {\bf 15} (1977) 480;
H.~Terazawa,
%``Subquark Model of Leptons and Quarks,''
Phys.\ Rev.\ D {\bf 22} (1980) 184.

\bibitem{composite-higgs}
S.~Dimopoulos, J.~Preskill,
%``Massless Composites With Massive Constituents,''
Nucl.\ Phys.\ B {\bf 199} (1982) 206;
D.~B.~Kaplan, H.~Georgi,
%``SU(2) x U(1) Breaking by Vacuum Misalignment,''
Phys.\ Lett.\ B {\bf 136} (1984) 183;
D.~B.~Kaplan, H.~Georgi, S.~Dimopoulos,
%``Composite Higgs Scalars,''
Phys.\ Lett.\ B {\bf 136} (1984) 187;
H.~Georgi, D.~B.~Kaplan, P.~Galison,
%``Calculation of the Composite Higgs Mass,''
Phys.\ Lett.\ B {\bf 143} (1984) 152;
T.~Banks,
%``CONSTRAINTS ON SU(2) x U(1) BREAKING BY VACUUM MISALIGNMENT,''
Nucl.\ Phys.\ B {\bf 243} (1984) 125;
H.~Georgi, D.~B.~Kaplan,
%``Composite Higgs and Custodial SU(2),''
Phys.\ Lett.\ B {\bf 145} (1984) 216;
M.~J.~Dugan, H.~Georgi, D.~B.~Kaplan,
%``Anatomy of a Composite Higgs Model,''
Nucl.\ Phys.\ B {\bf 254} (1985) 299;
H.~Georgi,
%``A Tool Kit for Builders of Composite Models,''
Nucl.\ Phys.\ B {\bf 266} (1986) 274.

\bibitem{Agashe:2004rs}
K.~Agashe, R.~Contino and A.~Pomarol,
%``The Minimal composite Higgs model,''
Nucl.\ Phys.\ B {\bf 719} (2005) 165
[hep-ph/0412089].

\bibitem{Bellazzini:2014yua}
B.~Bellazzini, C.~Csáki, J.~Serra,
%``Composite Higgses,''
Eur.\ Phys.\ J.\ C {\bf 74} (2014) 5,  2766
[arXiv:1401.2457 [hep-ph]].


%\cite{Sikivie:1980hm}
\bibitem{Sikivie:1980hm}
P.~Sikivie, L.~Susskind, M.~B.~Voloshin and V.~I.~Zakharov,
%``Isospin Breaking in Technicolor Models,''
Nucl.\ Phys.\ B {\bf 173} (1980) 189.

%\cite{Peskin:1991sw}
\bibitem{Peskin:1991sw}
M.~E.~Peskin and T.~Takeuchi,
%``Estimation of oblique electroweak corrections,''
Phys.\ Rev.\ D {\bf 46} (1992) 381.

\bibitem{Marandella:2005wd}
G.~Marandella, C.~Schappacher, A.~Strumia,
%``Little-Higgs corrections to precision data after LEP2,''
Phys.\ Rev.\ D {\bf 72} (2005) 035014
[hep-ph/0502096];
G.~Cacciapaglia, C.~Csaki, G.~Marandella, A.~Strumia,
%``The Minimal Set of Electroweak Precision Parameters,''
Phys.\ Rev.\ D {\bf 74} (2006) 033011
[hep-ph/0604111];
M.~Ciuchini, E.~Franco, S.~Mishima, L.~Silvestrini,
%``Electroweak Precision Observables, New Physics and the Nature of a 126 GeV Higgs Boson,''
JHEP {\bf 1308} (2013) 106
[arXiv:1306.4644 [hep-ph]].






\bibitem{EWPOCHM}
K.~Agashe, R.~Contino,
%``The Minimal composite Higgs model and electroweak precision tests,''
Nucl.\ Phys.\ B {\bf 742} (2006) 59
[hep-ph/0510164];
K.~Agashe, R.~Contino, L.~Da Rold, A.~Pomarol,
%``A Custodial symmetry for Zb anti-b,''
Phys.\ Lett.\ B {\bf 641} (2006) 62
[hep-ph/0605341];
G.~F.~Giudice, C.~Grojean, A.~Pomarol, R.~Rattazzi,
%``The Strongly-Interacting Light Higgs,''
JHEP {\bf 0706} (2007) 045
[hep-ph/0703164];
R.~Barbieri, B.~Bellazzini, V.~S.~Rychkov, A.~Varagnolo,
%``The Higgs boson from an extended symmetry,''
Phys.\ Rev.\ D {\bf 76} (2007) 115008
[arXiv:0706.0432 [hep-ph]];
P.~Lodone,
%``Vector-like quarks in a 'composite' Higgs model,''
JHEP {\bf 0812} (2008) 029
[arXiv:0806.1472 [hep-ph]];
M.~Gillioz,
%``A Light composite Higgs boson facing electroweak precision tests,''
Phys.\ Rev.\ D {\bf 80} (2009) 055003
[arXiv:0806.3450 [hep-ph]];
C.~Anastasiou, E.~Furlan, J.~Santiago,
%``Realistic Composite Higgs Models,''
Phys.\ Rev.\ D {\bf 79} (2009) 075003
[arXiv:0901.2117 [hep-ph]];
G.~Panico, A.~Wulzer,
%``The Discrete Composite Higgs Model,''
JHEP {\bf 1109} (2011) 135
[arXiv:1106.2719 [hep-ph]];
S.~De Curtis, M.~Redi, A.~Tesi,
%``The 4D Composite Higgs,''
JHEP {\bf 1204} (2012) 042
[arXiv:1110.1613 [hep-ph]];
D.~Marzocca, M.~Serone, J.~Shu,
%``General Composite Higgs Models,''
JHEP {\bf 1208} (2012) 013
[arXiv:1205.0770 [hep-ph]];
A.~Orgogozo, S.~Rychkov,
%``The S parameter for a Light Composite Higgs: a Dispersion Relation Approach,''
JHEP {\bf 1306} (2013) 014
[arXiv:1211.5543 [hep-ph]];
D.~Pappadopulo, A.~Thamm, R.~Torre,
%``A minimally tuned composite Higgs model from an extra dimension,''
JHEP {\bf 1307} (2013) 058
[arXiv:1303.3062 [hep-ph]];
C.~Grojean, O.~Matsedonskyi, G.~Panico,
%``Light top partners and precision physics,''
JHEP {\bf 1310} (2013) 160
[arXiv:1306.4655 [hep-ph]].

\bibitem{Frigerio:2011zg}
M.~Frigerio, J.~Serra, A.~Varagnolo,
%``Composite GUTs: models and expectations at the LHC,''
JHEP {\bf 1106} (2011) 029
[arXiv:1103.2997 [hep-ph]].

\bibitem{Pomarol:2008bh}%(ST+Z->bb+pp->tttt+composite top)
M.~Carena, E.~Ponton, J.~Santiago and C.~E.~M.~Wagner,
%``Light Kaluza Klein States in Randall-Sundrum Models with Custodial SU(2),''
Nucl.\ Phys.\ B {\bf 759} (2006) 202
[hep-ph/0607106];
A.~Pomarol, J.~Serra,
%``Top Quark Compositeness: Feasibility and Implications,''
Phys.\ Rev.\ D {\bf 78} (2008) 074026
[arXiv:0806.3247 [hep-ph]];
D.~Pappadopulo, A.~Thamm and R.~Torre,
%``A minimally tuned composite Higgs model from an extra dimension,''
JHEP {\bf 1307} (2013) 058
[arXiv:1303.3062 [hep-ph]].


\bibitem{Bellazzini:2012tv}
B.~Bellazzini, C.~Csaki, J.~Hubisz, J.~Serra, J.~Terning,
%``Composite Higgs Sketch,''
JHEP {\bf 1211} (2012) 003
[arXiv:1205.4032 [hep-ph]].


\bibitem{Azatov:2013ura}%(S+Higgs pheno)
M.~Gillioz, R.~Grober, C.~Grojean, M.~Muhlleitner, E.~Salvioni,
%``Higgs Low-Energy Theorem (and its corrections) in Composite Models,''
JHEP {\bf 1210} (2012) 004
[arXiv:1206.7120 [hep-ph]];
A.~Azatov, J.~Galloway,
%``Electroweak Symmetry Breaking and the Higgs Boson: Confronting Theories at Colliders,''
Int.\ J.\ Mod.\ Phys.\ A {\bf 28} (2013) 1330004
[arXiv:1212.1380];
A.~Falkowski, F.~Riva and A.~Urbano,
%``Higgs at last,''
JHEP {\bf 1311} (2013) 111
[arXiv:1303.1812 [hep-ph]];
A.~Azatov, R.~Contino, A.~Di Iura, J.~Galloway,
%``New Prospects for Higgs Compositeness in $h \to Z\gamma$,''
Phys.\ Rev.\ D {\bf 88} (2013) 7,  075019
[arXiv:1308.2676 [hep-ph]];
M.~Gillioz, R.~Gröber, A.~Kapuvari, M.~Mühlleitner,
%``Vector-like Bottom Quarks in Composite Higgs Models,''
JHEP {\bf 1403} (2014) 037
[arXiv:1311.4453 [hep-ph]].


\bibitem{Barbieri:2008zt}%(ST+Z->bb+flavour symmetry+collider signatures)
R.~Barbieri, G.~Isidori, D.~Pappadopulo,
%``Composite fermions in Electroweak Symmetry Breaking,''
JHEP {\bf 0902} (2009) 029
[arXiv:0811.2888 [hep-ph]];
O.~Matsedonskyi,
%``On Flavour and Naturalness of Composite Higgs Models,''
JHEP {\bf 1502} (2015) 154
[arXiv:1411.4638 [hep-ph]].

\bibitem{Barbieri:2012tu}%(FCNC-constraint+U(3)+U(2)+ST+Z->bb)
R.~Barbieri, D.~Buttazzo, F.~Sala, D.~M.~Straub, A.~Tesi,
%``A 125 GeV composite Higgs boson versus flavour and electroweak precision tests,''
JHEP {\bf 1305} (2013) 069
[arXiv:1211.5085 [hep-ph]].

%\cite{Csaki:2008zd}
\bibitem{Csaki:2008zd}%(FCNC-constraint+ST?+Z->bb?)
C.~Csaki, A.~Falkowski, A.~Weiler,
%``The Flavor of the Composite Pseudo-Goldstone Higgs,''
JHEP {\bf 0809} (2008) 008
[arXiv:0804.1954 [hep-ph]];
K.~Agashe, A.~Azatov, L.~Zhu,
%``Flavor Violation Tests of Warped/Composite SM in the Two-Site Approach,''
Phys.\ Rev.\ D {\bf 79} (2009) 056006
[arXiv:0810.1016 [hep-ph]].

\bibitem{Vignaroli:2012si}%(Z->bb+b->s\gamma)
N.~Vignaroli,
%``$\Delta$ F=1 constraints on composite Higgs models with LR parity,''
Phys.\ Rev.\ D {\bf 86} (2012) 115011
[arXiv:1204.0478 [hep-ph]].

\bibitem{Mrazek:2011iu}%(nonminimal-composite-Higgs+higgs-pheno)
B.~Gripaios, A.~Pomarol, F.~Riva, J.~Serra,
%``Beyond the Minimal Composite Higgs Model,''
JHEP {\bf 0904} (2009) 070
[arXiv:0902.1483 [hep-ph]];
J.~Mrazek, A.~Pomarol, R.~Rattazzi, M.~Redi, J.~Serra, A.~Wulzer,
%``The Other Natural Two Higgs Doublet Model,''
Nucl.\ Phys.\ B {\bf 853} (2011) 1
[arXiv:1105.5403 [hep-ph]];
M.~Redi, A.~Tesi,
%``Implications of a Light Higgs in Composite Models,''
JHEP {\bf 1210} (2012) 166
[arXiv:1205.0232 [hep-ph]];
E.~Bertuzzo, T.~S.~Ray, H.~de Sandes, C.~A.~Savoy,
%``On Composite Two Higgs Doublet Models,''
JHEP {\bf 1305} (2013) 153
[arXiv:1206.2623 [hep-ph]];
M.~Montull, F.~Riva,
%``Higgs discovery: the beginning or the end of natural EWSB?,''
JHEP {\bf 1211} (2012) 018
[arXiv:1207.1716 [hep-ph]];
M.~Chala,
%``$h \rightarrow \gamma\gamma$ excess and Dark Matter from Composite Higgs Models,''
JHEP {\bf 1301} (2013) 122
[arXiv:1210.6208 [hep-ph]].

\bibitem{Frigerio:2012uc}%(nonminimal-composite-Higgs+Higgs-pheno+DM)
M.~Frigerio, A.~Pomarol, F.~Riva, A.~Urbano,
%``Composite Scalar Dark Matter,''
 JHEP {\bf 1207} (2012) 015
 [arXiv:1204.2808 [hep-ph]].


\bibitem{Contino:2013kra}
R.~Contino, C.~Grojean, M.~Moretti, F.~Piccinini, R.~Rattazzi,
%``Strong Double Higgs Production at the LHC,''
JHEP {\bf 1005} (2010) 089
[arXiv:1002.1011 [hep-ph]];
I.~Low, A.~Vichi,
%``On the production of a composite Higgs boson,''
Phys.\ Rev.\ D {\bf 84} (2011) 045019
[arXiv:1010.2753 [hep-ph]];
R.~Contino, D.~Marzocca, D.~Pappadopulo, R.~Rattazzi,
%``On the effect of resonances in composite Higgs phenomenology,''
JHEP {\bf 1110} (2011) 081
[arXiv:1109.1570 [hep-ph]];
A.~Azatov, J.~Galloway,
%``Light Custodians and Higgs Physics in Composite Models,''
Phys.\ Rev.\ D {\bf 85} (2012) 055013
[arXiv:1110.5646 [hep-ph]];
R.~Contino, M.~Ghezzi, M.~Moretti, G.~Panico, F.~Piccinini, A.~Wulzer,
%``Anomalous Couplings in Double Higgs Production,''
JHEP {\bf 1208} (2012) 154
[arXiv:1205.5444 [hep-ph]];
R.~Contino, M.~Ghezzi, C.~Grojean, M.~Muhlleitner, M.~Spira,
%``Effective Lagrangian for a light Higgs-like scalar,''
JHEP {\bf 1307} (2013) 035
[arXiv:1303.3876 [hep-ph]];
C.~Delaunay, C.~Grojean, G.~Perez,
%``Modified Higgs Physics from Composite Light Flavors,''
JHEP {\bf 1309} (2013) 090
[arXiv:1303.5701 [hep-ph]];
A.~Banfi, A.~Martin, V.~Sanz,
%``Probing top-partners in Higgs+jets,''
JHEP {\bf 1408} (2014) 053
[arXiv:1308.4771 [hep-ph]];
M.~Montull, F.~Riva, E.~Salvioni, R.~Torre,
%``Higgs Couplings in Composite Models,''
Phys.\ Rev.\ D {\bf 88} (2013) 095006
[arXiv:1308.0559 [hep-ph]];
R.~Contino, C.~Grojean, D.~Pappadopulo, R.~Rattazzi, A.~Thamm,
%``Strong Higgs Interactions at a Linear Collider,''
JHEP {\bf 1402} (2014) 006
[arXiv:1309.7038 [hep-ph]];
T.~Flacke, J.~H.~Kim, S.~J.~Lee, S.~H.~Lim,
%``Constraints on composite quark partners from Higgs searches,''
JHEP {\bf 1405} (2014) 123
[arXiv:1312.5316 [hep-ph]];
C.~Grojean, E.~Salvioni, M.~Schlaffer, A.~Weiler,
%``Very boosted Higgs in gluon fusion,''
JHEP {\bf 1405} (2014) 022
[arXiv:1312.3317 [hep-ph]];
M.~Carena, L.~Da Rold, E.~Pontón,
%``Minimal Composite Higgs Models at the LHC,''
JHEP {\bf 1406} (2014) 159
[arXiv:1402.2987 [hep-ph]];
A.~Carmona, F.~Goertz,
%``A naturally light Higgs without light Top Partners,''
JHEP {\bf 1505} (2015) 002
[arXiv:1410.8555 [hep-ph]];
G.~Buchalla, O.~Cata, C.~Krause,
%``A Systematic Approach to the SILH Lagrangian,''
Nucl.\ Phys.\ B {\bf 894} (2015) 602
[arXiv:1412.6356 [hep-ph]].


\bibitem{Pomarol:2012qf}
A.~Pomarol, F.~Riva,
%``The Composite Higgs and Light Resonance Connection,''
JHEP {\bf 1208} (2012) 135
[arXiv:1205.6434 [hep-ph]];
O.~Matsedonskyi, G.~Panico, A.~Wulzer,
%``Light Top Partners for a Light Composite Higgs,''
JHEP {\bf 1301} (2013) 164
[arXiv:1204.6333 [hep-ph]].


\bibitem{Gherghetta:2004sq}%(gauge coupling unif in the CHM)
K.~Agashe, A.~Delgado, R.~Sundrum,
%``Grand unification in RS1,''
Annals Phys.\  {\bf 304} (2003) 145
[hep-ph/0212028];
T.~Gherghetta,
%``Partly supersymmetric grand unification,''
Phys.\ Rev.\ D {\bf 71} (2005) 065001
[hep-ph/0411090].

\bibitem{Barnard:2014tla}
 J.~Barnard, T.~Gherghetta, T.~S.~Ray, A.~Spray,
%``The Unnatural Composite Higgs,''
JHEP {\bf 1501} (2015) 067
[arXiv:1409.7391 [hep-ph]].

\bibitem{Asano:2014wra}
M.~Asano and R.~Kitano,
%``Partially Composite Dark Matter,''
JHEP {\bf 1409} (2014) 171
[arXiv:1406.6374 [hep-ph]].


\bibitem{Redi:2011zi}%(FCNC-constraint-10 TeV+flavour U(3) symmetry+LHC-pheno)
M.~Redi, A.~Weiler,
%``Flavor and CP Invariant Composite Higgs Models,''
JHEP {\bf 1111} (2011) 108
[arXiv:1106.6357 [hep-ph]].

\bibitem{Redi:2013pga}%(FCNC-constraint-lepton-u(2)-1 TeV+LHC pheno)
M.~Redi,
%``Leptons in Composite MFV,''
JHEP {\bf 1309} (2013) 060
[arXiv:1306.1525 [hep-ph]].

\bibitem{Delaunay:2013pwa}
K.~Agashe, A.~Belyaev, T.~Krupovnickas, G.~Perez, J.~Virzi,
%``LHC Signals from Warped Extra Dimensions,''
Phys.\ Rev.\ D {\bf 77} (2008) 015003
[hep-ph/0612015];
B.~Lillie, L.~Randall and L.~T.~Wang,
%``The Bulk RS KK-gluon at the LHC,''
JHEP {\bf 0709} (2007) 074
[hep-ph/0701166];
K.~Agashe, H.~Davoudiasl, S.~Gopalakrishna, T.~Han, G.~Y.~Huang, G.~Perez, Z.~G.~Si, A.~Soni,
%``LHC Signals for Warped Electroweak Neutral Gauge Bosons,''
Phys.\ Rev.\ D {\bf 76} (2007) 115015
[arXiv:0709.0007 [hep-ph]];
M.~Carena, A.~D.~Medina, B.~Panes, N.~R.~Shah, C.~E.~M.~Wagner,
%``Collider phenomenology of gauge-Higgs unification scenarios in warped extra dimensions,''
Phys.\ Rev.\ D {\bf 77} (2008) 076003
[arXiv:0712.0095 [hep-ph]];
R.~Contino, G.~Servant,
%``Discovering the top partners at the LHC using same-sign dilepton final states,''
JHEP {\bf 0806} (2008) 026
[arXiv:0801.1679 [hep-ph]];
K.~Agashe, S.~Gopalakrishna, T.~Han, G.~Y.~Huang, A.~Soni,
%``LHC Signals for Warped Electroweak Charged Gauge Bosons,''
Phys.\ Rev.\ D {\bf 80} (2009) 075007
[arXiv:0810.1497 [hep-ph]];
%J.~A.~Aguilar-Saavedra,
%%``Identifying top partners at LHC,''
%JHEP {\bf 0911} (2009) 030
%[arXiv:0907.3155 [hep-ph]];
J.~Mrazek, A.~Wulzer,
%``A Strong Sector at the LHC: Top Partners in Same-Sign Dileptons,''
Phys.\ Rev.\ D {\bf 81} (2010) 075006
[arXiv:0909.3977 [hep-ph]];
K.~Agashe, A.~Azatov, T.~Han, Y.~Li, Z.~G.~Si, L.~Zhu,
%``LHC Signals for Coset Electroweak Gauge Bosons in Warped/Composite PGB Higgs Models,''
Phys.\ Rev.\ D {\bf 81} (2010) 096002
[arXiv:0911.0059 [hep-ph]];
G.~Dissertori, E.~Furlan, F.~Moortgat, P.~Nef,
%``Discovery potential of top-partners in a realistic composite Higgs model with early LHC data,''
JHEP {\bf 1009} (2010) 019
[arXiv:1005.4414 [hep-ph]];
N.~Vignaroli,
%``Early discovery of top partners and test of the Higgs nature,''
Phys.\ Rev.\ D {\bf 86} (2012) 075017
[arXiv:1207.0830 [hep-ph]];
G.~Cacciapaglia, A.~Deandrea, L.~Panizzi, S.~Perries, V.~Sordini,
%``Heavy Vector-like quark with charge 5/3 at the LHC,''
JHEP {\bf 1303} (2013) 004
[arXiv:1211.4034 [hep-ph]];
A.~De Simone, O.~Matsedonskyi, R.~Rattazzi, A.~Wulzer,
%``A First Top Partner Hunter's Guide,''
JHEP {\bf 1304} (2013) 004
[arXiv:1211.5663 [hep-ph]];
%M.~Buchkremer, G.~Cacciapaglia, A.~Deandrea, L.~Panizzi,
%%``Model Independent Framework for Searches of Top Partners,''
%Nucl.\ Phys.\ B {\bf 876} (2013) 376
%[arXiv:1305.4172 [hep-ph]];
J.~Li, D.~Liu, J.~Shu,
%``Towards the fate of natural composite Higgs model through single $t^\prime$ search at the 8 TeV LHC,''
JHEP {\bf 1311} (2013) 047
[arXiv:1306.5841 [hep-ph]];
M.~Redi, V.~Sanz, M.~de Vries, A.~Weiler,
%``Strong Signatures of Right-Handed Compositeness,''
JHEP {\bf 1308} (2013) 008
[arXiv:1305.3818 [hep-ph]];
C.~Delaunay, T.~Flacke, J.~Gonzalez-Fraile, S.~J.~Lee, G.~Panico, G.~Perez,
%``Light Non-degenerate Composite Partners at the LHC,''
JHEP {\bf 1402} (2014) 055
[arXiv:1311.2072 [hep-ph]];
O.~Matsedonskyi, F.~Riva, T.~Vantalon,
%``Composite Charge 8/3 Resonances at the LHC,''
JHEP {\bf 1404} (2014) 059
[arXiv:1401.3740 [hep-ph]];
H.~C.~Cheng, J.~Gu,
%``Top seesaw with a custodial symmetry, and the 126 GeV Higgs,''
JHEP {\bf 1410} (2014) 002
[arXiv:1406.6689 [hep-ph]];
B.~Gripaios, T.~Müller, M.~A.~Parker, D.~Sutherland,
%``Search Strategies for Top Partners in Composite Higgs models,''
JHEP {\bf 1408} (2014) 171
[arXiv:1406.5957 [hep-ph]];
A.~Azatov, G.~Panico, G.~Perez, Y.~Soreq,
%``On the Flavor Structure of Natural Composite Higgs Models & Top Flavor Violation,''
JHEP {\bf 1412} (2014) 082
[arXiv:1408.4525 [hep-ph]];
M.~Backović, T.~Flacke, J.~H.~Kim, S.~J.~Lee,
%``Boosted Event Topologies from TeV Scale Light Quark Composite Partners,''
JHEP {\bf 1504} (2015) 082
[arXiv:1410.8131 [hep-ph]];
S.~Kanemura, K.~Kaneta, N.~Machida, T.~Shindou,
%``New resonance scale and fingerprint identification in minimal composite Higgs models,''
Phys.\ Rev.\ D {\bf 91} (2015) 11,  115016
[arXiv:1410.8413 [hep-ph]];
A.~Thamm, R.~Torre, A.~Wulzer,
%``Future tests of Higgs compositeness: direct vs indirect,''
arXiv:1502.01701 [hep-ph];
A.~Azatov, D.~Chowdhury, D.~Ghosh, T.~S.~Ray,
%``Same sign di-lepton candles of the composite gluons,''
arXiv:1505.01506 [hep-ph];
J.~Serra,
%``Beyond the Minimal Top Partner Decay,''
arXiv:1506.05110 [hep-ph].

\bibitem{Cacciapaglia:2014uja}
J.~Barnard, T.~Gherghetta, T.~S.~Ray,
%``UV descriptions of composite Higgs models without elementary scalars,''
JHEP {\bf 1402} (2014) 002
[arXiv:1311.6562 [hep-ph]];
G.~Ferretti, D.~Karateev,
%``Fermionic UV completions of Composite Higgs models,''
JHEP {\bf 1403} (2014) 077
[arXiv:1312.5330 [hep-ph]];
G.~Cacciapaglia, F.~Sannino,
%``Fundamental Composite (Goldstone) Higgs Dynamics,''
JHEP {\bf 1404} (2014) 111
[arXiv:1402.0233 [hep-ph]];
A.~Hietanen, R.~Lewis, C.~Pica, F.~Sannino,
%``Fundamental Composite Higgs Dynamics on the Lattice: SU(2) with Two Flavors,''
JHEP {\bf 1407} (2014) 116
[arXiv:1404.2794 [hep-lat]];
G.~Ferretti,
%``UV Completions of Partial Compositeness: The Case for a SU(4) Gauge Group,''
JHEP {\bf 1406} (2014) 142
[arXiv:1404.7137 [hep-ph]];
A.~Parolini,
%``Phenomenological aspects of supersymmetric composite Higgs models,''
Phys.\ Rev.\ D {\bf 90} (2014) 11,  115026
[arXiv:1405.4875 [hep-ph]];
M.~Geller, O.~Telem,
%``Holographic Twin Higgs Model,''
Phys.\ Rev.\ Lett.\  {\bf 114} (2015) 191801
[arXiv:1411.2974 [hep-ph]];
B.~Gripaios, M.~Nardecchia, S.~A.~Renner,
%``Composite leptoquarks and anomalies in $B$-meson decays,''
JHEP {\bf 1505} (2015) 006
[arXiv:1412.1791 [hep-ph]];
M.~Low, A.~Tesi, L.~T.~Wang,
%``Twin Higgs mechanism and a composite Higgs boson,''
Phys.\ Rev.\ D {\bf 91} (2015) 095012
[arXiv:1501.07890 [hep-ph]];
M.~Golterman, Y.~Shamir,
%``Top quark induced effective potential in a composite Higgs model,''
Phys.\ Rev.\ D {\bf 91} (2015) 094506
[arXiv:1502.00390 [hep-ph]].

\bibitem{Contino:2006nn}
R.~Contino, T.~Kramer, M.~Son, R.~Sundrum,
%``Warped/composite phenomenology simplified,''
JHEP {\bf 0705} (2007) 074
[hep-ph/0612180].

\bibitem{Kaplan:1991dc}
D.~B.~Kaplan,
%``Flavor at SSC energies: A New mechanism for dynamically generated fermion masses,''
Nucl.\ Phys.\ B {\bf 365} (1991) 259.

%\cite{Agashe:2004cp}
\bibitem{Agashe:2004cp}
K.~Agashe, G.~Perez, A.~Soni,
%``Flavor structure of warped extra dimension models,''
Phys.\ Rev.\ D {\bf 71} (2005) 016002
[hep-ph/0408134].

\bibitem{Blanke:2008zb}%(FCNC-constraint-20 TeV)
M.~Blanke, A.~J.~Buras, B.~Duling, S.~Gori, A.~Weiler,
%``$\Delta$ F=2 Observables and Fine-Tuning in a Warped Extra Dimension with Custodial Protection,''
JHEP {\bf 0903} (2009) 001
[arXiv:0809.1073 [hep-ph]];
O.~Gedalia, G.~Isidori, G.~Perez,
%``Combining Direct & Indirect Kaon CP Violation to Constrain the Warped KK Scale,''
Phys.\ Lett.\ B {\bf 682} (2009) 200
[arXiv:0905.3264 [hep-ph]].

\bibitem{Barbieri:2012uh}%(FCNC-constraint-quark-lepton+U(3)+U(2)-3 TeV)
R.~Barbieri, D.~Buttazzo, F.~Sala, D.~M.~Straub,
%``Flavour physics from an approximate $U(2)^3$ symmetry,''
JHEP {\bf 1207} (2012) 181
[arXiv:1203.4218 [hep-ph]].



\bibitem{Agashe:2006iy}%(FCNC-constraint-lepton-10 TeV)
K.~Agashe, A.~E.~Blechman, F.~Petriello,
%``Probing the Randall-Sundrum geometric origin of flavor with lepton flavor violation,''
Phys.\ Rev.\ D {\bf 74} (2006) 053011
[hep-ph/0606021];
C.~Csaki, Y.~Grossman, P.~Tanedo, Y.~Tsai,
%``Warped penguin diagrams,''
Phys.\ Rev.\ D {\bf 83} (2011) 073002
[arXiv:1004.2037 [hep-ph]].


\bibitem{Csaki:2008qq}%(FCNC-constraint-lepton+neutrino masses&mixing-A4-no-stringent-bounds-on-KK??)
C.~Csaki, C.~Delaunay, C.~Grojean, Y.~Grossman,
%``A Model of Lepton Masses from a Warped Extra Dimension,''
 JHEP {\bf 0810} (2008) 055
 [arXiv:0806.0356 [hep-ph]];
F.~del Aguila, A.~Carmona, J.~Santiago,
%``Neutrino Masses from an A4 Symmetry in Holographic Composite Higgs Models,''
JHEP {\bf 1008} (2010) 127
[arXiv:1001.5151 [hep-ph]].

\bibitem{Cacciapaglia:2007fw}%(FCNC-suppression+U(3))
G.~Cacciapaglia, C.~Csaki, J.~Galloway, G.~Marandella, J.~Terning, A.~Weiler,
%``A GIM Mechanism from Extra Dimensions,''
JHEP {\bf 0804} (2008) 006
[arXiv:0709.1714 [hep-ph]];
M.~Redi,
%``Composite MFV and Beyond,''
Eur.\ Phys.\ J.\ C {\bf 72} (2012) 2030
[arXiv:1203.4220 [hep-ph]];
M.~König, M.~Neubert, D.~M.~Straub,
%``Dipole operator constraints on composite Higgs models,''
Eur.\ Phys.\ J.\ C {\bf 74} (2014) 7,  2945
[arXiv:1403.2756 [hep-ph]].

%\cite{Nevzorov:2015sha}
\bibitem{Nevzorov:2015sha}
R.~Nevzorov, A.~W.~Thomas,
%``$E_6$ inspired composite Higgs model,''
Phys.\ Rev.\ D {\bf 92} (2015) 075007 [arXiv:1507.02101 [hep-ph]].


%\cite{e6ssm}
\bibitem{e6ssm}
%D.~Suematsu, Y.~Yamagishi,
%%``Radiative symmetry breaking in a supersymmetric model with an extra U(1),''
%Int.\ J.\ Mod.\ Phys.\  A {\bf 10} (1995) 4521;
%E.~Ma,
%%``Neutrino masses in an extended gauge model with E(6) particle content,''
%Phys.\ Lett.\  B {\bf 380} (1996) 286;
%E.~Keith, E.~Ma,
%%``Efficacious Extra U(1) Factor for the Supersymmetric Standard Model,''
%Phys.\ Rev.\  D {\bf 54} (1996) 3587;
%E.~Keith, E.~Ma,
%%``Generic consequences of a supersymmetric U(1) gauge factor at the TeV
%%scale,''
%Phys.\ Rev.\  D {\bf 56} (1997) 7155;
%D.~Suematsu,
%%``Neutralino decay in the mu problem solvable extra U(1) models,''
%Phys.\ Rev.\  D {\bf 57} (1998) 1738;
Y.~Daikoku, D.~Suematsu,
%``Mass bound of the lightest neutral Higgs scalar in the extra U(1)
%models,''
Phys.\ Rev.\  D {\bf 62} (2000) 095006;
T.~Hambye, E.~Ma, M.~Raidal, U.~Sarkar,
%``Allowable low-energy E(6) subgroups from leptogenesis,''
Phys.\ Lett.\  B {\bf 512} (2001) 373;
S.~F.~King, S.~Moretti, R.~Nevzorov,
%``Theory and phenomenology of an exceptional supersymmetric standard
%model,''
Phys.\ Rev.\  D {\bf 73} (2006) 035009;
S.~F.~King, S.~Moretti, R.~Nevzorov,
%``Exceptional supersymmetric standard model,''
Phys.\ Lett.\  B {\bf 634} (2006) 278;
S.~F.~King, S.~Moretti, R.~Nevzorov,
%``Spectrum of Higgs particles in the ESSM,''
arXiv:hep-ph/0601269;
S.~F.~King, S.~Moretti, R.~Nevzorov,
%``E(6)SSM,''
AIP Conf.\ Proc.\  {\bf 881} (2007) 138;
%[arXiv:hep-ph/0610002]
S.~F.~King, S.~Moretti, R.~Nevzorov,
%``Gauge coupling unification in the exceptional supersymmetric standard
%model,''
Phys.\ Lett.\  B {\bf 650} (2007) 57;
%[arXiv:hep-ph/0701064];
R.~Howl, S.~F.~King,
%``Minimal E(6) Supersymmetric Standard Model,''
JHEP {\bf 0801} (2008) 030;
%[arXiv:0708.1451 [hep-ph]];
P.~Athron, S.~F.~King, D.~J.~Miller, S.~Moretti, R.~Nevzorov,
%``The Constrained E$_6$SSM,''
arXiv:0810.0617 [hep-ph];
S.~Hesselbach, D.~J.~Miller, G.~Moortgat-Pick, R.~Nevzorov and M.~Trusov,
%``Theoretical upper bound on the mass of the LSP in the MNSSM,''
Phys.\ Lett.\ B {\bf 662} (2008) 199;
%[arXiv:0712.2001 [hep-ph]].
P.~Athron, S.~F.~King, D.~J.~Miller, S.~Moretti, R.~Nevzorov,
%``Predictions of the Constrained Exceptional Supersymmetric Standard Model,''
Phys.\ Lett.\  B {\bf 681} (2009) 448;
%[arXiv:0901.1192 [hep-ph]];
P.~Athron, S.~F.~King, D.~J.~Miller, S.~Moretti, R.~Nevzorov,
%``The Constrained Exceptional Supersymmetric Standard Model,''
Phys.\ Rev.\  D {\bf 80} (2009) 035009;
%[arXiv:0904.2169 [hep-ph]];
J.~P.~Hall, S.~F.~King, R.~Nevzorov, S.~Pakvasa, M.~Sher,
%``Novel Higgs Decays and Dark Matter in the E(6)SSM,''
Phys.\ Rev.\  D {\bf 83} (2011) 075013;
%[arXiv:1012.5114 [hep-ph]];
P.~Athron, S.~F.~King, D.~J.~Miller, S.~Moretti, R.~Nevzorov,
%``LHC Signatures of the Constrained Exceptional Supersymmetric Standard
%Model,''
Phys.\ Rev.\ D {\bf 84} (2011) 055006;
%[arXiv:1102.4363 [hep-ph]];
P.~Athron, D.~Stockinger,  A.~Voigt,
%``Threshold Corrections in the Exceptional Supersymmetric Standard Model,''
Phys.\ Rev.\ D {\bf 86} (2012) 095012;
%[arXiv:1209.1470 [hep-ph]].
P.~Athron, S.~F.~King, D.~J.~Miller, S.~Moretti, R.~Nevzorov,
%``Constrained Exceptional Supersymmetric Standard Model with a Higgs Near 125 GeV,''
Phys.\ Rev.\ D {\bf 86} (2012) 095003;
%[arXiv:1206.5028 [hep-ph]].
R.~Nevzorov,
%``$E_6$ inspired supersymmetric models with exact custodial symmetry,''
Phys.\ Rev.\ D {\bf 87} (2013) 015029;
%[arXiv:1205.5967 [hep-ph]].
P.~Athron, M.~Binjonaid, S.~F.~King,
%``Fine Tuning in the Constrained Exceptional Supersymmetric Standard Model,''
Phys.\ Rev.\ D {\bf 87} (2013) 115023;
%[arXiv:1302.5291 [hep-ph]].
D.~J.~Miller, A.~P.~Morais, P.~N.~Pandita,
%``Constraining Grand Unification using first and second generation sfermions,''
Phys.\ Rev.\ D {\bf 87} (2013) 015007;
%[arXiv:1208.5906 [hep-ph]];
M.~Sperling, D.~Stöckinger, A.~Voigt,
%``Renormalization of vacuum expectation values in spontaneously broken gauge theories,''
JHEP {\bf 1307} (2013) 132;
%[arXiv:1305.1548 [hep-ph]];
R.~Nevzorov,
%``Quasifixed point scenarios and the Higgs mass in the E6 inspired supersymmetric models,''
Phys.\ Rev.\ D {\bf 89} (2014) 5,  055010;
%[arXiv:1309.4738 [hep-ph]].
R.~Nevzorov, S.~Pakvasa,
%``Exotic Higgs decays in the $E_6$ inspired SUSY models,''
Phys.\ Lett.\ B {\bf 728} (2014) 210;
%[arXiv:1308.1021 [hep-ph]].
M.~Sperling, D.~Stöckinger, A.~Voigt,
%``Renormalization of vacuum expectation values in spontaneously broken gauge theories: Two-loop results,''
JHEP {\bf 1401}, 068 (2014);
%[arXiv:1310.7629 [hep-ph]].
P.~Athron, M.~Mühlleitner, R.~Nevzorov, A.~G.~Williams,
%``Non-Standard Higgs Decays in U(1) Extensions of the MSSM,''
JHEP {\bf 1501} (2015) 153;
%[arXiv:1410.6288 [hep-ph]].
R.~Nevzorov, S.~Pakvasa,
%``Nonstandard Higgs decays in the E6 inspired SUSY models,''
arXiv:1411.0386 [hep-ph];
R.~Nevzorov,
%``LHC Signatures and Cosmological Implications of the E$_6$ Inspired SUSY Models,''
PoS EPS {\bf -HEP2015} (2015) 381
[arXiv:1510.05387 [hep-ph]];
P.~Athron, D.~Harries, R.~Nevzorov and A.~G.~Williams,
%``$E_6$ Inspired SUSY benchmarks, dark matter relic density and a 125 GeV Higgs,''
Phys.\ Lett.\ B {\bf 760} (2016) 19
[arXiv:1512.07040 [hep-ph]];
S.~F.~King and R.~Nevzorov,
%``750 GeV Diphoton Resonance from Singlets in an Exceptional Supersymmetric Standard Model,''
JHEP {\bf 1603} (2016) 139
[arXiv:1601.07242 [hep-ph]];
P.~Athron, M.~Muhlleitner, R.~Nevzorov and A.~G.~Williams,
%``Exotic Higgs decays in U(1) extensions of the MSSM,''
arXiv:1602.04453 [hep-ph];
P.~Athron, D.~Harries, R.~Nevzorov and A.~G.~Williams,
%``Dark matter in a constrained E$_{6}$ inspired SUSY model,''
JHEP {\bf 1612} (2016) 128 [arXiv:1610.03374 [hep-ph]];
P.~Athron, A.~W.~Thomas, S.~J.~Underwood and M.~J.~White,
%``Dark matter candidates in the constrained Exceptional Supersymmetric Standard Model,''
Phys.\ Rev.\ D {\bf 95} (2017)  035023 [arXiv:1611.05966 [hep-ph]].

%\cite{King:2008qb}
\bibitem{King:2008qb}
S.~F.~King, R.~Luo, D.~J.~Miller and R.~Nevzorov,
%``Leptogenesis in the Exceptional Supersymmetric Standard Model: Flavour dependent lepton asymmetries,''
JHEP {\bf 0812} (2008) 042
[arXiv:0806.0330 [hep-ph]].



\bibitem{Phillips:2014fgb}
K.~S.~Babu {\it et al.},
%``Working Group Report: Baryon Number Violation,''
arXiv:1311.5285 [hep-ph];
D.~G.~Phillips, II {\it et al.},
%``Neutron-Antineutron Oscillations: Theoretical Status and Experimental Prospects,''
Phys.\ Rept.\  {\bf 612} (2016) 1
[arXiv:1410.1100 [hep-ex]].

\bibitem{Kronfeld:2013uoa}
A.~S.~Kronfeld {\it et al.},
%``Project X: Physics Opportunities,''
arXiv:1306.5009 [hep-ex].

\bibitem{Agashe:2005vg}
K.~Agashe, R.~Contino and R.~Sundrum,
%``Top compositeness and precision unification,''
Phys.\ Rev.\ Lett.\  {\bf 95} (2005) 171804
[hep-ph/0502222].

\bibitem{Nevzorov:2016fxp}
R.~Nevzorov and A.~W.~Thomas,
%``Diphoton signature of neutral pseudo-Goldstone boson in the E6CHM at the CERN LHC,''
arXiv:1605.07313 [hep-ph];
R.~Nevzorov and A.~W.~Thomas,
%``E6 inspired composite Higgs model and 750 GeV diphoton excess,''
EPJ Web Conf.\  {\bf 125} (2016) 02021
[arXiv:1608.00320 [hep-ph]].

\bibitem{CPasym-SM}
M.~A.~Luty,
%``Baryogenesis Via Leptogenesis,''
Phys.\ Rev.\  D {\bf 45} (1992) 455;
M.~Flanz, E.~A.~Paschos and U.~Sarkar,
%``Baryogenesis from a lepton asymmetric universe,''
Phys.\ Lett.\  B {\bf 345} (1995) 248
[Erratum-ibid.\  B {\bf 382} (1996) 447];
M.~Plumacher,
%``Baryogenesis and lepton number violation,''
Z.\ Phys.\  C {\bf 74} (1997) 549;
W.~Buchmuller and M.~Plumacher,
%``CP asymmetry in Majorana neutrino decays,''
Phys.\ Lett.\  B {\bf 431} (1998) 354.

\bibitem{Davidson:2008bu}
S.~Davidson, E.~Nardi and Y.~Nir,
%``Leptogenesis,''
Phys.\ Rept.\  {\bf 466} (2008) 105
[arXiv:0802.2962 [hep-ph]].

\bibitem{Kuzmin:1985mm}
V.~A.~Kuzmin, V.~A.~Rubakov and M.~E.~Shaposhnikov,
%``On The Anomalous Electroweak Baryon Number Nonconservation In The Early
%universe,''
Phys.\ Lett.\  B {\bf 155} (1985) 36;
V.~A.~Rubakov and M.~E.~Shaposhnikov,
%Electroweak baryon number non-conservation in the \uppercase{E}arly
%  \uppercase{U}niverse and in high-energy collisions.
Usp.\ Fiz.\ Nauk, {\bf 166} (1996) 493.

\end{thebibliography}
\end{document}